\documentclass[9pt,twocolumn,twoside]{pnas-new}
 \templatetype{pnasresearcharticle}
 \usepackage{amsmath,amssymb}
 \usepackage{stmaryrd}
 \usepackage{latexsym}
 \usepackage{CJK}
 \usepackage{graphicx}
 \usepackage{indentfirst}
 \usepackage{listings}
 \usepackage{xcolor}
 \usepackage{booktabs}
 \usepackage{stmaryrd}
 \usepackage{multirow}
 \usepackage{verbatim}
 \usepackage{makecell}
 \usepackage{longtable}
 \usepackage{hyperref}

 \hypersetup{colorlinks,breaklinks,
            linkcolor=blue,urlcolor=blue,anchorcolor=blue,citecolor=blue}

\definecolor{ao(english)}{rgb}{0.0, 0.5, 0.0}

\newcolumntype{C}[1]{>{\centering\let\newline\\\arraybackslash\hspace{0pt}}m{#1}}

\title{Grain-Boundary Topological Phase Transitions}

\author[a]{Kongtao Chen}
\author[b]{David J. Srolovitz}
\author[b,1]{Jian Han} 

\affil[a]{Department of Materials Science and Engineering, University of Pennsylvania, Philadelphia, PA 19104 USA}
\affil[b]{Department of Materials Science and Engineering, City University of Hong Kong, Kowloon, Hong Kong SAR, China}

\date{\today}

\leadauthor{Chen} 

\significancestatement{
We reveal the existence of a topological phase transition (Kosterlitz-Thouless type) in grain boundaries (GBs) -- important internal surfaces in crystalline materials. 
GB dynamics are controlled by the formation/migration of line defects (disconnections) with dislocation and step character. 
Below the GB KT transition, disconnections of opposite signs are bound as pairs, while above it, they unbind and proliferate. 
We demonstrate that GB KT transitions provide a  fundamental understanding of many GB properties, the temporal evolution of microstructure and how it deforms/degrades under stress. 
}

\authorcontributions{All authors  designed the research and wrote the paper; K.C. performed simulations and analyzed data.}
\authordeclaration{The authors declare no conflict of interest.}
\correspondingauthor{\textsuperscript{1}To whom correspondence should be addressed. E-mail: jianhan@cityu.edu.hk}

\keywords{materials science $|$ thermodynamics $|$ phase transition $|$ grain boundary $|$ grain growth} 

\begin{abstract}
The formation and migration of disconnections (line defects constrained to the grain boundary (GB) plane with both dislocation and step character) control many of the kinetic and dynamical properties of GBs and the polycrystalline materials of which they are central constituents. 
We demonstrate that GBs undergo a finite-temperature topological phase transition of the Kosterlitz-Thouless (KT) type. 
This phase transition corresponds to the screening of long-range interactions between (and unbinding of) disconnections.
This phase transition leads to abrupt change in the behavior of GB migration, GB sliding, and roughening. 
We analyze this KT transition through mean-field theory, renormalization group theory, and kinetic Monte Carlo simulations, and examine how this transition affects microstructure-scale phenomena such as grain growth stagnation, abnormal grain growth and superplasticity.
\end{abstract}

\begin{document}

\maketitle
\thispagestyle{firststyle}
\ifthenelse{\boolean{shortarticle}}{\ifthenelse{\boolean{singlecolumn}}{\abscontentformatted}{\abscontent}}{}


\dropcap{P}hysical properties of polycrystalline materials depend strongly on the properties of their constituent grain boundaries (GBs).
Like most material properties, GB properties are functions of temperature and may change abruptly at temperatures corresponding to phase transitions.
Such GB phase transitions may explain the existence of critical temperatures at which abrupt changes in the nature of several physical phenomena, including grain growth stagnation \cite{Holm2010} and superplasticity \cite{Edington1976}.  

There are several types of GB phase transitions discussed in the literature. These include thermodynamic phase transitions such as GB structural transitions or faceting/defaceting transitions (which are first-order) \cite{Cantwell2014,Frolov2013,Meiners2020}, roughening transitions (divergence in the height-height correlation function) which may be continuous \cite{Rottman1986}, and improper transitions where the GB transforms from solid-like to glass-like \cite{Zhang2009}. 
In this paper, we discuss a new class of fundamentally different GB phase transitions.
We identify a GB topological 
phase transition of a type of the class originally discussed by Kosterlitiz and Thouless \cite{Kosterlitz1973}.
Such topological transitions may be thought of as defect binding/unbinding transitions.
The most important type of defects for GB dynamics are disconnections~\cite{Han2018}. 
Disconnections are line defects, constrained to lie within the GB 
and characterized by both Burgers vector $\mathbf{b}$ and step height $h$.  
The set of possible disconnection modes $\{\mathbf{b},h\}$ is set by  GB bicrystallography.
Disconnections (like dislocations) are topological defects, as seen through a Burgers circuit analysis \cite{Han2018}.

Below the topological or Kosterlitz-Thouless (KT) transition temperature $T_{\text{KT}}$, the interaction between disconnections is long-range, decaying as the inverse of their separation. 
The formation and migration of disconnections are severely restricted and GB mobility tends to be small (with important exceptions).
On the other hand, above $T_{\text{KT}}$, the long-range elastic field of disconnections is effectively screened.
Hence, the KT transition may be viewed as a screening (or sliding) transition, where the screening parameter (diaelastic constant) diverges at $T_{\text{KT}}$.
The KT transition leads to abrupt changes in the GB migration mobility, roughness, sliding coefficient, etc.

While the KT transition leads to GB roughening, this transition fundamentally differs from the  roughening transition widely discussed in the literature \cite{Rottman1986}. 
While this classic roughening phase transition is topological, the steps have no long-range elastic interactions and are not topological defects (nonlocal imperfections corresponding to singularities in an order parameter characterizing a broken symmetry \cite{Irvine2013}).
On the other hand, the dislocation nature of disconnections implies that disconnections are fundamentally topological defects.
Hence, the KT transition discussed in this paper corresponds to the screening of long-range elastic interactions, while the classical roughening transition arises from topologically stable configurations of steps (without long-range interactions).

We analyze the KT transition first through mean-field theory, then apply renormalization group analysis to accurately predict the main features of this transition by formal extension to the thermodynamic limit (i.e., infinite length scales). 
The results are confirmed through a series of kinetic Monte Carlo (kMC) simulations.
Our analysis shows that the Kosterlitz-Thouless (KT) transition temperature depends on the driving force for GB motion.
For example, in curvature-driven grain growth, we find that at a fixed temperature,  large grains are more likely to be below $T_{\text{KT}}$ (low mobility) and small grains above it (high mobility).
This is a possible explanation of why grain growth in pure materials often stagnates at large grain size and superplasticity is generally restricted to small grain sizes and high temperatures. 
We confirm these results by comparing our renormalization group prediction of the grain size at which grain growth stagnation occurs with simulation data from the literature \cite{Holm2010}.

\section*{Disconnection Topological Transitions\label{RGsection}}

We describe the migration of GBs in terms of the motion of disconnections. Motion of disconnections leads to both GB migration (step motion) and shear across the GB (dislocation motion). We demonstrate that an abrupt change in GB behavior may result from an abrupt change in how disconnections interact; this is a topological phase transition.

We first consider the case of a dislocation in a two-dimensional (2d) elastic medium ($x$-$y$ plane) interacting with dislocation dipoles, following the general approach originally described by Kosterlitz and Thouless \cite{Kosterlitz1973}. 
Here, we assume that the Burgers vectors are parallel to the $x$-axis: $\mathbf{b} = b\hat{\mathbf{x}}$.
The elastic interaction energy of a test dislocation and another dislocation varies as the logarithm of their separation and the interaction force is minus the gradient of this energy with respect to test dislocation displacement  (i.e., decaying as the inverse of their separation). 
When multiple dislocations are present, the total force on the test dislocation is simply the sum of the forces from each of these.
The divergence of the force on the test dislocation (of unit $\mathbf{b}$) at $\mathbf{r}$, $\mathbf{f}(\mathbf{r})$ is  proportional to the Burgers vector distribution around the test dislocation $\rho(\mathbf{r})$:  
$\nabla \cdot\langle\mathbf{f}\rangle = 4\pi K \rho(\mathbf{r})$, where $\rho(\mathbf{r})$ is the Burgers vector density and the  constant $K \equiv \mu/[4\pi(1 - \nu)]$ ($\mu$ and $\nu$ are the elastic shear modulus and Poisson ratio). 
(Angular brackets  $\langle...\rangle$  denote the time average of a fluctuating quantity.)
 Dislocation dipoles in the material polarize (dislocations in the dipole separate) under the action of a (Peach-Koehler) force \cite{Sutton1995}. 
The  (polarized) dipole moment is $\int \langle \mathbf{p} \rangle \mathrm{d}\mathbf{r} = \int \mathbf{r} \rho(\mathbf{r}) \mathrm{d} \mathbf{r}$, where $\mathbf{p}$ is the instantaneous dipole moment density. 
The  distribution of polarized dipoles exert a force on the test dislocation:  $\langle \mathbf{f}_p \rangle = -4\pi K\langle\mathbf{p}\rangle$.  
The total force on the test dislocation is the sum of the applied force $ \mathbf{f}$ and that associated with the induced dipoles: $\langle\mathbf{f}_\text{t}\rangle = \mathbf{f} + \langle\mathbf{f}_p\rangle = \mathbf{f} - 4\pi K\langle\mathbf{p}\rangle$. 
The dipole moment is induced by the total force, $\langle\mathbf{p}\rangle = \boldsymbol{\chi} \langle\mathbf{f}_\text{t}\rangle $ (to leading order in the force),
where, in analogy to electrostatics, we define the susceptibility (tensor) as $\boldsymbol{\chi} \equiv (\partial\langle\mathbf{p}\rangle/\partial\langle\mathbf{f}_\text{t}\rangle)\rvert_{\langle\mathbf{f}_\text{t}\rangle = \mathbf{0}}$. 

The total force on the test dislocation can then be expressed as the external force {\it screened} by the induced dipoles  $\langle\mathbf{f}_\text{t}\rangle = \mathbf{f}/\epsilon$, where $\epsilon$ is the diaelastic constant (akin to the dielectric constant in electrostatics). 
The diaelastic constant describes the  strength of the screening of the applied force  by the induced dislocation dipoles:  $\epsilon\mathbf{I} = \mathbf{I} + 4\pi K \boldsymbol{\chi}$ ($\mathbf{I}$ is the identity matrix).

\subsection*{Disconnections on Grain Boundaries}
We now apply this approach to disconnections on a nominally flat GB which, in this 2d model, is the line $y = 0$.  
In this case, the Burgers vector density is $\rho(x, y) = \rho(x)/\delta$,  where $\delta$ is the disconnection core size. 
The external applied force on the test dislocation (with unit $\mathbf{b}$) is equal to the applied (shear) stress $\tau$ parallel to $\mathbf{b}$,  $f =\tau$. 
In this case, $\epsilon = 1 + 4\pi K\chi$, where $\chi$ is the the $xx$ component of the susceptibility tensor.
The distribution of the dislocation dipole moment associated with all dipoles with separation smaller than $r$ is $\langle p(r,x)\rangle
= \int_\delta^r \langle b r^\prime\rangle n(r', x) \,\mathrm{d} r'$, 
where $\langle b r^\prime\rangle$ is the moment of a dipole of separation $r'$ and $n(r')$ is the number of dipoles with separation in $[r', r'+\mathrm{d} r']$ per unit length between $x$ and $x+\mathrm{d} x$.   
Since we assume that the GB is uniform, we can drop the explicit dependence on $x$ and write the susceptibility as  $\chi(r) = (\partial\langle p\rangle/\partial f)\rvert_{f=0} = \int_{\delta}^r \alpha(r') n(r') \,\mathrm{d} r'$
($\alpha(r') \equiv (\partial \langle br'\rangle/\partial \tau)\rvert_{\tau=0}$ is the dipole polarizability).
The diaelastic constant, then, becomes  
\begin{equation}\label{epsilonr}
\epsilon(r) = 1 + 4\pi K\int_{\delta}^r \alpha(r') n(r') \,\mathrm{d} r'. 
\end{equation}

We evaluate the dipole polarizability and dipole number density $n(r)$ by assuming that the dipoles are in thermal equilibrium (Maxwell-Boltzmann distribution) as in the Debye-H\"uckel approximation for charged fluids.
The polarizability is
\begin{equation}\label{alphar}
\alpha(r) 
= \left.\frac{\partial}{\partial\tau} 
\dfrac{\sum_{b'=\pm b} b'r e^{-\beta(E_\text{c} - \tau b' r)} }
{\sum_{b'=\pm b} e^{-\beta(E_\text{c} - \tau b' r)} } 
\right|_{\tau=0}
= \beta b^2 r^2,
\end{equation}
where $E_\text{c}$ is the disconnection core energy (per unit length), $w$ is the width of the system in the direction parallel to the disconnection line and $\beta \equiv w/k_\text{B} T$.
The number density of dipoles of separation $[r,r+\mathrm{d} r]$ is
\begin{equation}\label{nr}
n(r) 
= \delta^{-3} e^{-\beta\left[2E_\text{c} + V(r) - (\psi h + \tau b)r\right]}, 
\end{equation}
where $\psi$ is the chemical potential jump across the GB \cite{Janssens2006}, and the elastic interaction energy (per unit length) of  the two disconnections in a dipole (separation $r$) is 
\begin{equation}\label{Vr}
V(r) = 2Kb^2 \int_\delta^r 
\frac{\mathrm{d} r'}{r' \epsilon(r')}. 
\end{equation}

This completes the derivation except for the determination of the dialectic constant $\epsilon(r)$, which can be determined through the self-consistent solution of \eqref{epsilonr} - \eqref{Vr}. 
It is useful to define the following dimensionless (reduced) quantities: the reduced length $l \equiv \ln\left({r}/{\delta}\right)$, 
the reduced inverse diaelastic constant $g \equiv \beta Kb^2/\epsilon(r)$, and the reduced dipole density $f \equiv \sqrt{r^3 n(r)}$.
These substitutions reduce four coupled equations to just two: 
\begin{equation}\label{RGequations}
\left\{
\def\arraystretch{2.2}
\begin{array}{l}
\displaystyle{ \frac{\mathrm{d} g^{-1}}{\mathrm{d} l} = 4\pi f^2 }\\
\displaystyle{ \frac{\mathrm{d} \ln f}{\mathrm{d} l}
= \frac{3}{2} - g + \frac{1}{2}\beta (\psi h + \tau b)r } 
\end{array}\right..
\end{equation}
In the limit that $r\rightarrow\delta$ (i.e., $l=0$ ), we find that $g(0) = \beta Kb^2$ and $\ln f(0) =  \beta[ (\psi h + \tau b)\delta-2E_\text{c}]/2$.

\subsection*{Topological Phase Transition}

Here, we examine the topological phase transition that occurs for disconnections on a GB that is associated with the screening of the disconnection fields. 
We do this first via a simple mean-field analysis (designed to provide qualitative, physical understanding) and then via a more rigorous renormalization group approach. 

\begin{figure}
\centering
\includegraphics[height=0.7\linewidth]{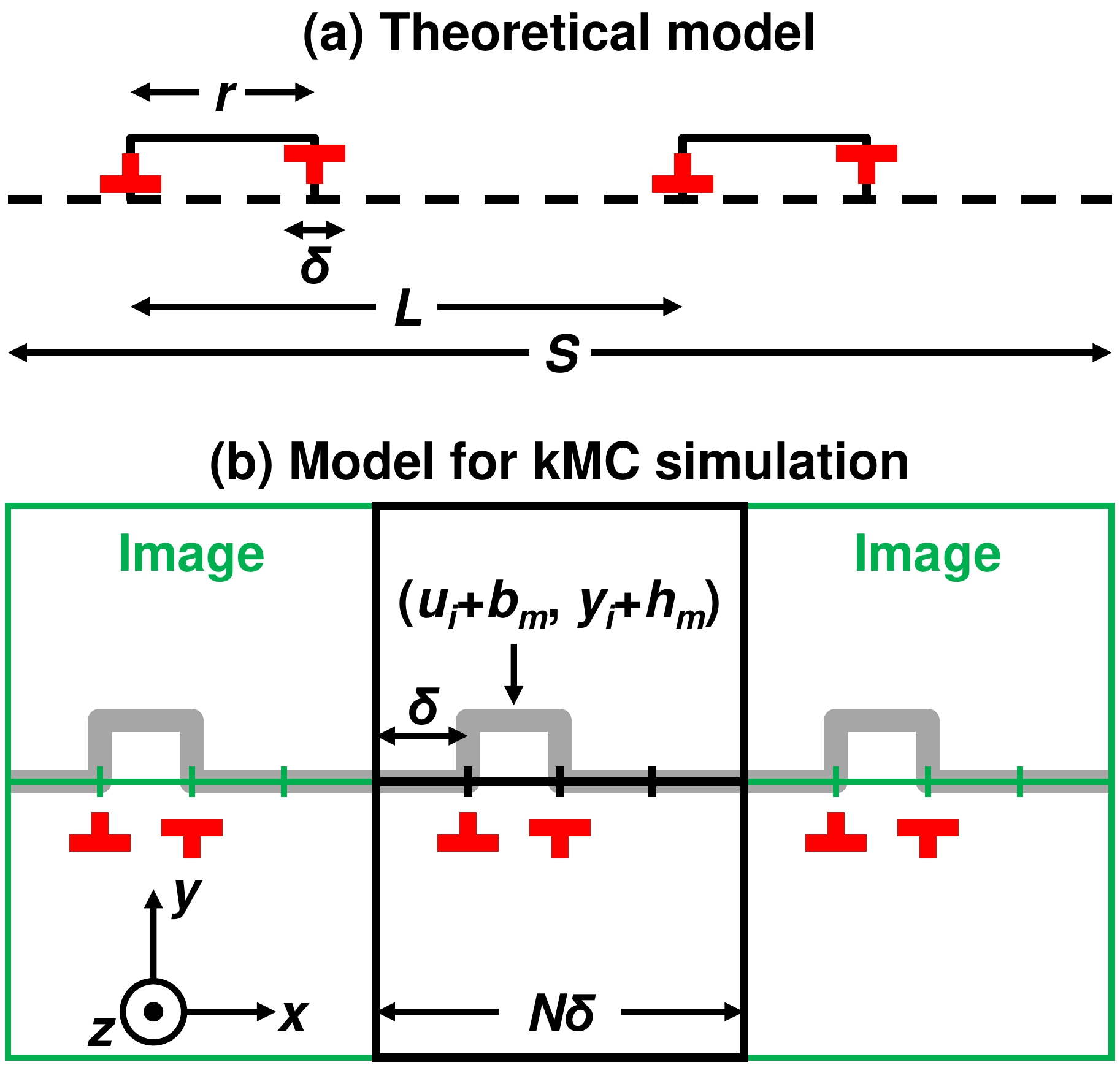}
\caption{\label{KMCmodel}
(a) Disconnection dipoles on a 1d GB in 2d. 
$\delta$, $r$, $L$ and $S$ denote the disconnection core size, the  separation of the disconnections in a  disconnection dipole, the average distance between the disconnection dipoles and the system size in the $x$-direction, respectively.
(b) The states of the system at $i$ before and after disconnection dipole nucleation are $(u_i, y_i)$ and $(u_i+b_m, y_i+h_m)$, respectively, for a pair of disconnections of mode $m$ at location $i$.
}
\end{figure}

As above, consider a bicrystal in 2d containing a flat 1d GB, as depicted in Fig.~\ref{KMCmodel};  the tilt axis is in $z$ and the nominal GB normal is in  $y$. 
We focus on a thermally equilibrated GB; incorporating the formation and annihilation of disconnection dipoles (Fig.~\ref{KMCmodel}a). 
The separation between the two disconnections in a dipole is $r$, the average distance between disconnection dipoles is $L$, and the size (length) of the GB is $S$. 
Since this model does not describe  GB structure on the atomic-scale, it does not account for such  phenomena  as premelting.

For simplicity, we assume that there is only one type of disconnection dipole on the GB, i.e., disconnection mode $(\pm b, \pm h)$.
If $b=0$ (pure-step mode), the free energy change associated with disconnection dipole formation on a flat GB is $F = 2E_\text{c} - k_\text{B} T \ln(S/\delta) < 0$ as $S\rightarrow\infty$, since the disconnection core energy $E_c$ is independent of GB size $S$\cite{Thomas2017,Han2018,Chen2019}  and the configurational entropy is proportional to $\ln (S/\delta)$.
This suggests that an infinitely large, 1d GB is rough at all $T>0$ K (i.e., the roughening transition temperature is $0$ K). 
If the GB is of finite size or where the GB is 2d, the roughening temperature is finite \cite{Swendsen1977}.
If $b\ne 0$, however, both entropy and disconnection elastic energy are proportional to $\ln (S/\delta)$; this suggests that there will be a critical temperature above which the entropy term dominates the free energy such that the roughening transition temperature is finite (in all dimensions). 

The effect of non-zero $b$ can be understood as follows. 
At low $T$, disconnections exist as closely bound dipoles; while at high $T$ the ratio of the separation between disconnections in a dipole to the spacing between dipoles is no longer small, such that the disconnection dipoles mix or, alternatively, the dipoles are unbound. 
Based on this idea, we can distinguish the low-$T$ from high-$T$ regimes based on whether $\langle r^2/L^2\rangle  \ll 1$ or $\gg 1$, respectively.  

The energy of a disconnection dipole has the form $U(r) = 2E_\text{c} + V(r)$, where the elastic potential energy is $ V(r) = 2Kb^2\ln(r/\delta)$. 
We first assume that the equilibrium disconnection dipole density  $1/L$ is low; i.e., $E_\text{c} \gg k_\text{B} T$.
The ensemble average (square of the) disconnection separation in   dipoles is  
\begin{equation}\label{rsquare}
\langle r^2 \rangle
= \dfrac{\int_\delta^\infty r^2 e^{-\beta U(r)} \mathrm{d} r}
{\int_\delta^\infty e^{-\beta U(r)} \mathrm{d} r}
= \delta^2 \left(\frac{2\beta Kb^2 - 1}{2\beta Kb^2 - 3}\right). 
\end{equation}
The average number of dipoles in length $L$ can be obtained by the grand-canonical ensemble average: 
\begin{equation*}
\langle \mathcal{N} \rangle
= \dfrac{\sum_{\mathcal{N}=0}^\infty \mathcal{N} \mathcal{Z}^\mathcal{N} \mathcal{P}^\mathcal{N}}
{\sum_{\mathcal{N}=0}^\infty \mathcal{Z}^\mathcal{N} \mathcal{P}^\mathcal{N}}
= \mathcal{Z} \mathcal{P} + \mathcal{O}(\mathcal{Z}^2), 
\end{equation*}
where $\mathcal{Z} \equiv e^{-2\beta E_\text{c}}$ and 
\begin{equation*}
\mathcal{P} \equiv \frac{1}{\delta^2} \int_0^L \mathrm{d} x \int_\delta^\infty  e^{-\beta V(r)}\mathrm{d} r
= \frac{L/\delta}{2\beta Kb^2 - 1}. 
\end{equation*}
The average disconnection dipole density $\langle 1/L\rangle$ is obtained by setting $\langle \mathcal{N}\rangle = 1$:  
\begin{equation}\label{oneoverL}
\left\langle\frac{1}{L}\right\rangle
= \left(\frac{1}{\delta}\right)\frac{ e^{-2\beta E_\text{c}}}{2\beta Kb^2 - 1}. 
\end{equation}
From \eqref{rsquare} and \eqref{oneoverL}, we find
\begin{equation}\label{rtoL}
\left\langle \frac{r^2}{L^2}\right\rangle
=
\dfrac{e^{-4\beta E_\text{c}}}
{
(2\beta Kb^2 - 3)(2\beta Kb^2 - 1)
}.
\end{equation}
This demonstrates that a critical temperature $T_\text{KT}$ for which  $\langle r^2/L^2 \rangle \to \infty$: 
\begin{equation}\label{kttc0}
T_{\text{KT}} = 2Kb^2w/3k_\text{B}.
\end{equation}
$T_{\text{KT}}$  is  the Kosterlitz-Thouless transition temperature.
For $T<T_{\text{KT}}$, the disconnections are bound pairs; i.e., disconnections exist as dipoles bound together by elastic interactions.
However, for $T>T_{\text{KT}}$ the disconnections are unbound; i.e., each disconnection is free to move independently - not bound to any other disconnection.

The GB roughness, as characterized by the standard deviation of the GB profile $\sigma_y$ scales as $\langle r^2/L^2\rangle^\frac{1}{4}$ and  diverges at $T_{\text{KT}}$ (in 1D GB  $T_\text{KT}$ is also the roughening transition temperature). 
This is discussed in more detail in SI Appendix, demonstrated in our kMC simulations (below), and observed in MD simulations \cite{Olmsted2007a}.

The mean-field theory reveals that the long-range elastic interaction between disconnections may result in the disconnection binding-unbinding (or pairing-unpairing) transition. 
However,  mean-field analysis rarely provides accurate predictions of the phenomenon near a phase transition (and thus fails to predict $T_\text{KT}$ accurately). 
This problem can often be overcome by application of renormalization group methods.
Following the spirit of renormalization group approaches for dislocations \cite{Kosterlitz1973,Khantha1994}, we look for numerical solutions of \eqref{RGequations}  to obtain $f(l)$ and $g(l)$. 

\begin{figure}
\centering
\includegraphics[height=0.5\linewidth]{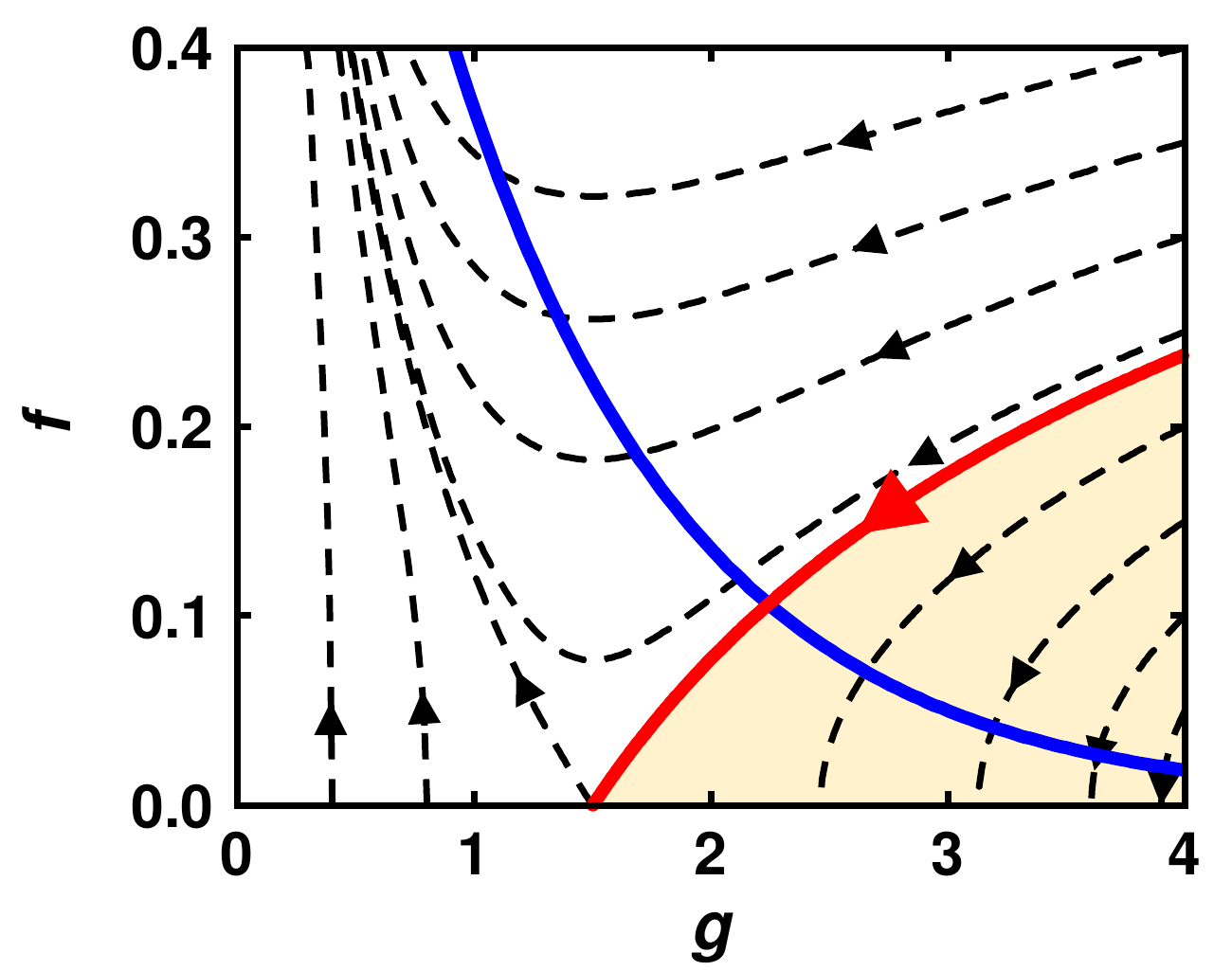}
\caption{\label{RGPhase}
Renormalization flows obtained by numerical solution of \eqref{RGequations} with $\psi=\tau = 0$ and different initial conditions $(g(0),f(0))$.
The arrows denote directions of increasing length scale $l$  (coarse-graining).
The red curve is the critical manifold.   
The flows in the shaded region converge to $(g>1.5, f=0)$ as $l \to \infty$, while the flows in the unshaded region converge to $(g=0, f \to\infty)$. 
The blue curve corresponds to $(g(0,T), f(0,T))$ where the temperature $T$ increases from bottom right to top left; the material and GB parameters are those used for the one-mode kMC simulation.
The Kosterlitz-Thouless transition temperature corresponds to the point where the red and blue curves cross.
}
\end{figure}

Any set of physical parameters/driving forces ($K$, $b$, $h$, $E_\text{c}$, $T$, $\psi$ and $\tau$) correspond to different initial conditions $(g(0),f(0))$ in the solution of \eqref{RGequations}; 
starting from each particular initial condition there is a flow in $(g(l), f(l))$  as $l$ varies from $0$ to $\infty$ (since $l$ is a length scale, this is coarse-graining that takes the system to the thermodynamic limit). 
This is depicted in Fig.~\ref{RGPhase} for different $(g(0),f(0))$.
Figure~\ref{RGPhase} shows that there are two types of fixed points as $l\to\infty$: (i) a ``superfluid phase'' $(g = 0, f \to \infty)$, where  screening  diminishes dislocation interactions (i.e., $\epsilon\to\infty$) such that there are many unbound disconnection (i.e., $r^3n \to \infty$) and the GB is rough, and (ii) an ``insulating phase'' $(g > 1.5, f = 0)$, where  screening  is limited ($\epsilon < 2Kb^2w/3k_\text{B} T$ as $r\to\infty$),  few unbound disconnections exist ($r^3n\rightarrow 0$) and the  GB is flat.
There is a critical manifold (red curve) in Fig.~\ref{RGPhase}; flows above the critical manifold converge to type (i) fixed points (unbound disconnection/rough GB phase), while flows below this manifold converge to type (ii) fixed points (bound disconnection/flat GB phase).

The KT transition temperature can be determined numerically. 
For a particular GB, $\big(g(l=0,T), f(l=0,T)\big)$ is a parametric curve;  temperature $T$ is the parameter (the blue curve in Fig.~\ref{RGPhase}). 
The point where  the red and blue curves cross (Fig.~\ref{RGPhase}) corresponds to $T_\text{KT}$. 
Formally, this temperature can also be determined from the condition that $f$ is scale-invariant:
\begin{equation}\label{kttc}
T_\text{KT} = \left[2Kb^2\epsilon^{-1}(r_\text{c}) - (\psi h + \tau b)r_\text{c}\right]w/3k_\text{B}, 
\end{equation}
where $r_\text{c}$ is the average disconnection separation  in the dipoles at $T_\text{KT}$  ($\epsilon(r_\text{c})$ must  be determined numerically). 
\eqref{kttc} is consistent with the  mean-field result \eqref{kttc0} ($\epsilon(r_\text{c}) = 1$) in the sparse disconnection ($E_c\gg Kb^2w$) and small driving force limits.
When multiple disconnection modes are present, $T_\text{KT}$ is dominated by the disconnection mode with the smallest $T_\text{KT}$.

Since the KT transition is associated with the screening of the long-range elastic interactions between disconnections and since disconnection motion is the underlying mechanism of GB migration, the KT transition should lead to a discontinuity in the temperature dependence of the GB mobility.
When $T<T_\text{KT}$, the activation energy for GB migration $Q$ includes both the disconnection glide barrier $E^*$ and the large scale barrier associated with elastic interactions \cite{Han2018,Chen2019,Chen2020,Chen2020a}; when $T>T_\text{KT}$, the elastic barrier is effectively screened. 
Hence, increasing $T$ through $T_\text{KT}$ leads to an abrupt decrease in the activation energy for GB migration $Q$; the slope of the GB mobility versus temperature curve should increase abruptly upon heating through $T_\text{KT}$. 
Such an abrupt increase in the  GB mobility slope versus $T$ is observed at  $T_\text{KT}$ in the kMC simulations shown below. 

Similar results were observed in the molecular dynamics (MD) simulations.
Homer et al.~\cite{Homer2014} observed that the slope of the mobility versus temperature of the  Ni $\Sigma 39$ $[111]$ $(752)$ symmetrical tilt GB changed abruptly at a finite $T$. 
For this GB,  $b = a_0/\sqrt{26}$ and $h = 3a_0/\sqrt{78}$ ($a_0$ is the lattice constant)\cite{Han2018}. 
With this input and the GB energy $\gamma \approx 0.5$~J/m$^2$, thickness $w=7.5a_0$, $K=9$ GPa (assuming $r_\text{c} \sim w/2$ is the largest disconnection distance in this periodic cell), we find that $\epsilon=7.6$ from Fig.~\ref{RGPhase}. 
and (\eqref{kttc}) $T_\text{KT} \approx 800$~K. 
So, the theoretical prediction of the temperature where an abrupt change of the activation energy for GB mobility is about $800$~K, which is close to the MD result under $\psi=0.025$~eV.~\cite{Homer2014}
The MD results of Homer et al.~\cite{Homer2014} also showed that the activation energy for GB migration $Q$ is an approximately linear function of the ``roughening temperature'', $T_\text{KT}$.  
We\cite{Chen2020a} previously showed that the activation energy for GB migration $Q$ varies linearly with $Kb^2$ and \eqref{kttc}  shows that  $Kb^2$ is proportional to $T_\text{KT}$; hence,  $Q$ is a linear function of  $T_\text{KT}$ as observed in MD. 
Based on their MD simulations, Olmsted et al.~\cite{Olmsted2007a} observed that at low temperatures and small driving forces, GBs migrate in a start-stop fashion, while at high temperatures/large driving forces, GBs migrate continuously.
This may be understood by noting that  below  the KT transition ($T<T_\text{KT}$, which is driving force dependent), disconnection nucleation barriers are high thus disconnection nucleation time is much longer than migration time, while above the transition,  nucleation (barriers) times are relatively short (disconnection screening effect) and are comparable with/smaller than migration times\cite{Chen2020a}.

\section*{Kinetic Monte Carlo Simulations}


Here, we compare the theoretical analysis with the results of disconnection-based kinetic Monte Carlo (kMC) simulations. 
Figure~\ref{KMCmodel}b shows the basic model employed in our kinetic Monte Carlo (kMC) simulations (the GB tilt axis and nominal GB normal are in $z$ and $y$ and the system is periodic in $x$). 
The GB is discretized into $N$ lattice sites along $x$. 
The state of GB site $i$ ($1\le i \le N$) is denoted by $(u_i(t), y_i(t))$, where $u_i$ is the relative (tangential) displacement of the upper grain with respect to the lower one (in $x$) and $y_i$ is the position (in $y$) of the GB at site $i$. 
Formation of a pair of disconnections of mode $m$ $(\pm b_m, \pm h_m)$ at  $i$ corresponds to  $(u_i, y_i) \to (u_i+b_m, y_i+h_m)$, as illustrated in Fig.~\ref{KMCmodel}b. 
See   \textit{Methods}  for a detailed description of the kMC algorithm and the definition of the dimensionless quantities  in this section. 
We report kMC results for two simulation cases:  
(i) a pure step mode, $h=1$ and
(ii) a single disconnection mode, $b=1$ and $h=1$.

For the parameters used in the kMC simulations, the renormalization group analysis (Fig.~\ref{RGPhase}) predicts  $T_\text{KT}=0.1$ for the  single disconnection mode kMC  and $T_\text{KT}=0$ for the pure-step simulations.
The roughening transition ($\sigma_y\to\infty$ and spatial correlation length $\to\infty$) and screening/sliding transition ($\epsilon\rightarrow\infty$)  occur at the same temperature  $T_\text{KT}$ for the case of a GB in a 2d bicrystal.  

\begin{figure*}
\centering
\includegraphics[height=0.8\linewidth]{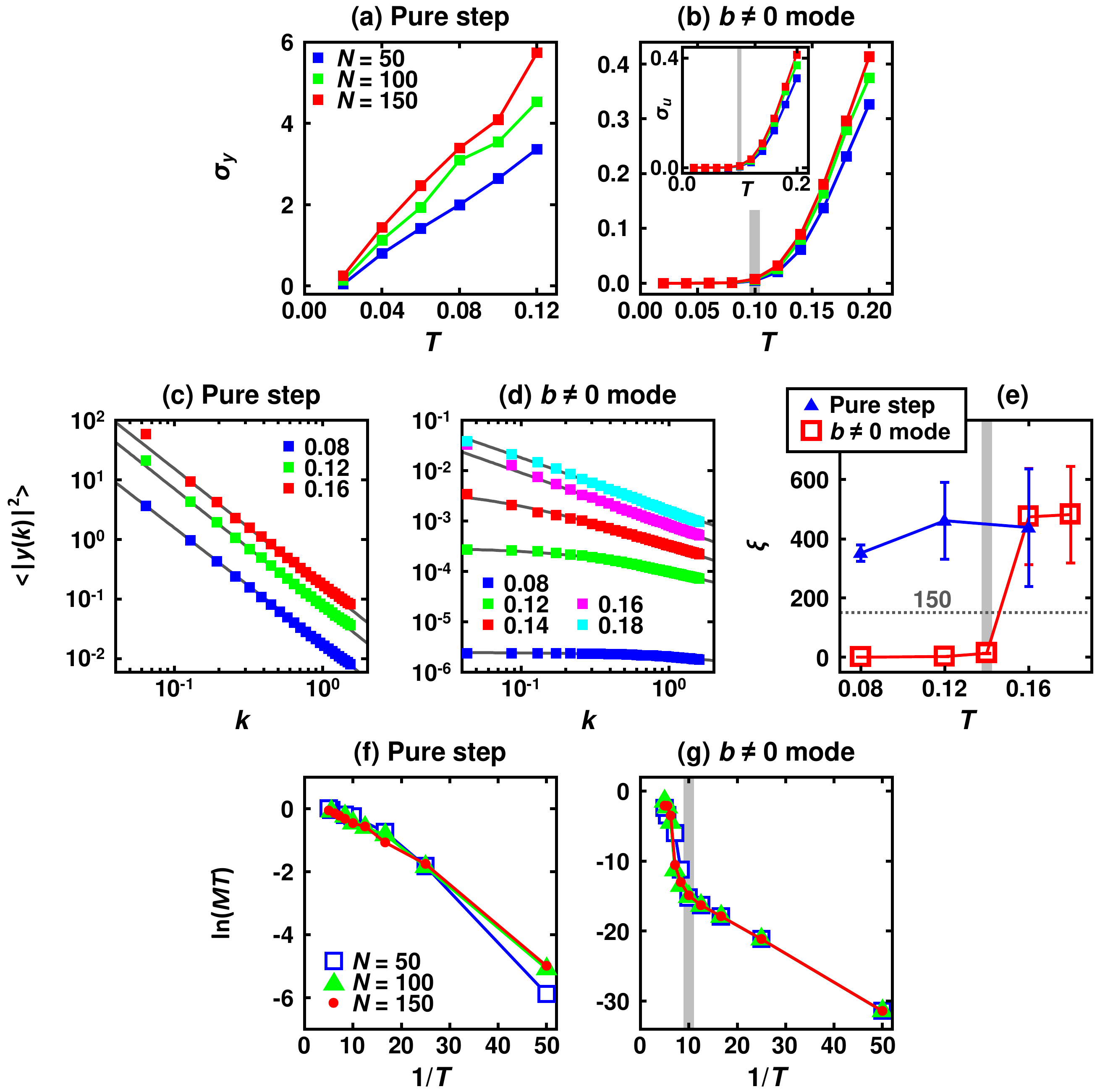}
\caption{\label{kmcresults} 
(a)-(b) show the GB roughness $\sigma_y$ vs. $T$ for the  (a) pure step and (b) single mode with $b\ne 0$; the insets in (b) is the standard deviations of shear $\sigma_u$ vs. temperature $T$. 
(c)-(d) show the average of the (square of the) magnitude of the Fourier transform ($k$ is a wave vector) of the equilibrium GB profile $y(x)$, $\langle \lvert y(k) \rvert^2\rangle$, for several temperatures (see the legend)  for the  (c) pure step and (d) single mode with $b\ne 0$.  
(e) shows the correlation length $\xi(T)$ (obtained from fitting $\xi=AT/(k^d + \xi^{-d})$ for each temperature to the kMC data in  (c)-(d);  the horizontal dashed line (at $\xi = 150$) is the kMC simulation cell period (in $x$). 
(f)-(g) show $\ln (MT)$ vs. $1/T$~\cite{Chen2020}, where $M$ is the GB mobility for the  (f) pure step and (g) single mode cases. 
The vertical gray lines label $T = 0.1$ in (b), and (g), and $T = 0.14$ in (e). 
}
\end{figure*}

Figures~\ref{kmcresults}a,b show the standard deviations of the GB profile $\sigma_y = \sqrt{\langle y^2 \rangle - \langle y \rangle^2}$, obtained from the kMC simulations. 
We recall that the roughening transition occurs at $T=0$ for pure steps in 2d; this is consistent with  Fig.~\ref{kmcresults}a for which the roughness varies smoothly with temperature across the entire simulation temperature range. 
On the other hand, introduction of a finite disconnection $\mathbf{b}$ (see Fig.~\ref{kmcresults}b) effectively suppresses roughening at low $T$ ($ \lesssim T_\text{KT}$). 
At low $T$ the roughness is nearly size-independent; this suggests the presence of very short-range correlations in the GB profile at low $T$. 
(The spatial correlation length $\xi$ is the length scale over which the two-point correlation between the heights of different points on the surface decays with their separation.)
Above $T_\text{KT}$, $\sigma_y$ increases rapidly with $T$ and a  strong size effect (larger roughening in larger systems) is observed. 
The presence of the near linear dependence of roughness on temperature and a strong size dependence above $T_\text{KT}$ is reminiscent of the roughening behavior in the pure step case (Fig.~\ref{kmcresults}a) at $T>0$. 
These are signatures of a finite-$T$ transition. 

In  finite-$\mathbf{b}$ systems, the  standard deviation of the shear $\sigma_u = \sqrt{\langle u^2 \rangle - \langle u \rangle^2}$ (see the insets in Figs.~\ref{kmcresults}b) show similar behavior as the GB profile roughening. 
The abrupt change in ``shear roughening'' suggests that shear roughening is also a characteristic of the disconnection  KT transition $T_\text{KT}$.
(In 3d, the GB profile and shear roughening need not occur at the same $T$.)

Equilibrium fluctuations in the GB profile provide direct evidence of the GB roughening transition. 
We expand the GB profile in a Fourier series, $y(x,t) = \sum_k y(k,t) e^{ikx}$ and measure the  equilibrium static GB profile spectrum  $\langle |y(k)|^2 \rangle$, where $\langle \cdot \rangle$ represents a time average. 
Liao et al.~\cite{Liao2018}  demonstrated that for pure steps, this spectrum should be described by
\begin{equation}\label{spectrumk2}
\langle |y(k)|^2 \rangle 
= T / [N\Gamma(k^2 + \xi^{-2})], 
\end{equation}
where $\xi$ is the correlation length and $\Gamma$ is the dimensionless GB stiffness. 
For disconnection with non-zero $\mathbf{b}$~\cite{Karma2012}, 
\begin{equation}\label{spectrumk1}
\langle |y(k)|^2 \rangle  = T/[N \mathcal{B}^2(k^1 + \xi^{-1})], 
\end{equation} 
where $\mathcal{B} \equiv b/h$ is the shear coupling factor. 
A correlation length $\xi$ is introduced here as a wavelength cutoff~\cite{Liao2018}. 
The KT transition theory suggests that $\xi\to\infty$ for $T>T_\text{KT}$~\cite{Liao2018}.

Figures~\ref{kmcresults}c-d show the spectra obtained from the kMC simulations. 
These results indeed demonstrate that $\langle |y(k)|^2 \rangle \propto T$, consistent with \eqref{spectrumk2} and \eqref{spectrumk1}. 
The kMC data for each temperature were fitted to the function $AT/(k^{d} + \xi^{-d})$, where $A$, $d$ and the correlation length $\xi$ are the fitting parameters ($d$ and $\xi$ are functions of $T$).
For  $b=0$ (Fig.~\ref{kmcresults}c), $d \approx 2$, while when $b\ne 0$ (Figs.~\ref{kmcresults}d), $d \approx 1$; consistent with Eqs. \ref{spectrumk2} and \ref{spectrumk1}.
The correlation length $\xi$ obtained by the fitting at each temperature is shown in Fig.~\ref{kmcresults}e. 
Since our kMC simulation were performed using a finite width GB  ($N=150$), we consider the GB  roughened when $\xi > 150$.
(Since $\xi$  diverges above  $T_\text{KT}$, it is not possible to obtain accurate measurements of  $\xi$ above $T_\text{KT}$.)
Using this operational definition, we find  that when $b=0$ the GB is rough at all temperatures, but  only rough at $T \ge 0.14\approx T_\text{KT}$ for  $b\ne 0$.
The small difference between the theoretical prediction ($T_\text{KT} = 0.1$) and the simulation result ($T_\text{KT} = 0.14$) may be  attributable to the finite GB width in the simulations and approximations in \eqref{spectrumk2} and \eqref{spectrumk1}.

The GB mobility may be related\cite{Trautt2006} to  fluctuations in the mean GB position $\bar{y}$: 
$M = N\bar{y}^2(\Delta t)/2 \Delta t T$, where $\Delta t$ is the time interval used in the calculation of the time correlation $\bar{y}$.
Figures~\ref{kmcresults}f,  g show the GB mobilities versus temperature from the kMC simulations. 
When the operative disconnection mode is a pure step mode  (Fig.~\ref{kmcresults}f), the GB mobility behaves in a quasi-Arrhenius fashion;  $\ln(MT) \propto -Q/T$~\cite{Chen2020}, where the activation energy $Q$ (i.e.,  slope of $\ln(MT)$ vs. $1/T$) is roughly temperature-independent 
(note that $Q$  may exhibit a weak temperature dependence in cases where the disconnection nucleation  and migration times are comparable~\cite{Chen2020a}). 
When $b\ne 0$ (Fig.~\ref{kmcresults}g), the activation energy $Q$ (slope) changes abruptly at $T\approx T_\text{KT}$. 
This is because at $T > T_\text{KT}$, the disconnection elastic fields are effectively screened such that the elastic contribution to the activation energy for disconnection formation is zero. 
The  temperature at which the activation energy for  mobility changes ($b\ne 0$) coincides  with   an abrupt change in both $\sigma_y$ and/or $\sigma_u$  (Fig.~\ref{kmcresults}b), i.e., $T_\text{KT}$.

The kMC simulations demonstrate that, when the activated disconnection mode has non-zero $\mathbf{b}$, a finite temperature dynamic phase transition occurs in the GB  (provided melting does not occur first).
Examination of the standard deviations of the GB profile $\sigma_y$ and the equilibrium GB fluctuation spectrum $\langle |y(k)|^2\rangle$ suggests that such a phase transition corresponds to the GB roughening transition.
The simultaneous transitions in the behavior of the standard deviations of the GB shear $\sigma_x$, the divergence of correlation length $\xi$ above critical temperature and the temperature dependence of GB mobility suggest that the roughening transition is a Kosterlitz-Thouless, topological phase transition. 
The kMC simulation results suggest that the abrupt changes in the temperature dependencies of $\sigma_y$, $\sigma_x$, $\langle |y(k)|^2\rangle$ and $M$ provides clear  evidence of a transition temperature for GB dynamics with $b \ne 0$ disconnections 
(see Fig.~\ref{kmcresults}) at a temperature consistent  with the KT transition temperature $T_\text{KT}$ predicted by the renormalization group theory, $T=0.1$ (section \textit{Disconnection Topological Transitions}).
In other words,  finite $b$ disconnection-mediated KT transitions can result in both GB roughening and changes in GB migration behavior.

\section*{Grain Growth Stagnation}

Grain growth stagnation is widely observed in both experiments~\cite{Barmak2006} and MD simulations~\cite{Holm2010}.
Holm and Foiles suggested that this stagnation is associated with the GB roughening transition~\cite{Holm2010}.
Here, we argue that this behavior is better described in terms of the GB KT transition. 

The driving force for grain growth is the reduction of the energy of the GB network in a polycrystal. 
In classical analyses of normal grain growth, we assume the GB energy is isotropic and  GB migration is overdamped.
This means that the GB velocity is proportional to its mean curvature $H$ (i.e.,  mean curvature flow).
The chemical potential jump across the GB is $\psi = \gamma H$ and the mean curvature scales (on average) as the inverse of the grain size, $D$;   $\psi$  decreases as $D$ increases.  
\eqref{kttc} shows that decreasing $\psi$ (increasing  grain size $D$) implies an increasing KT transition temperature, $T_\text{KT}(D)$.  
Therefore, during isothermal grain growth, the increase in the mean grain size results in fewer and fewer mobile grains (i.e., those with $T_\text{KT}(D)<T$). 
This may lead to grain growth stagnation.

\begin{figure}
\centering
\includegraphics[height=0.5\linewidth]{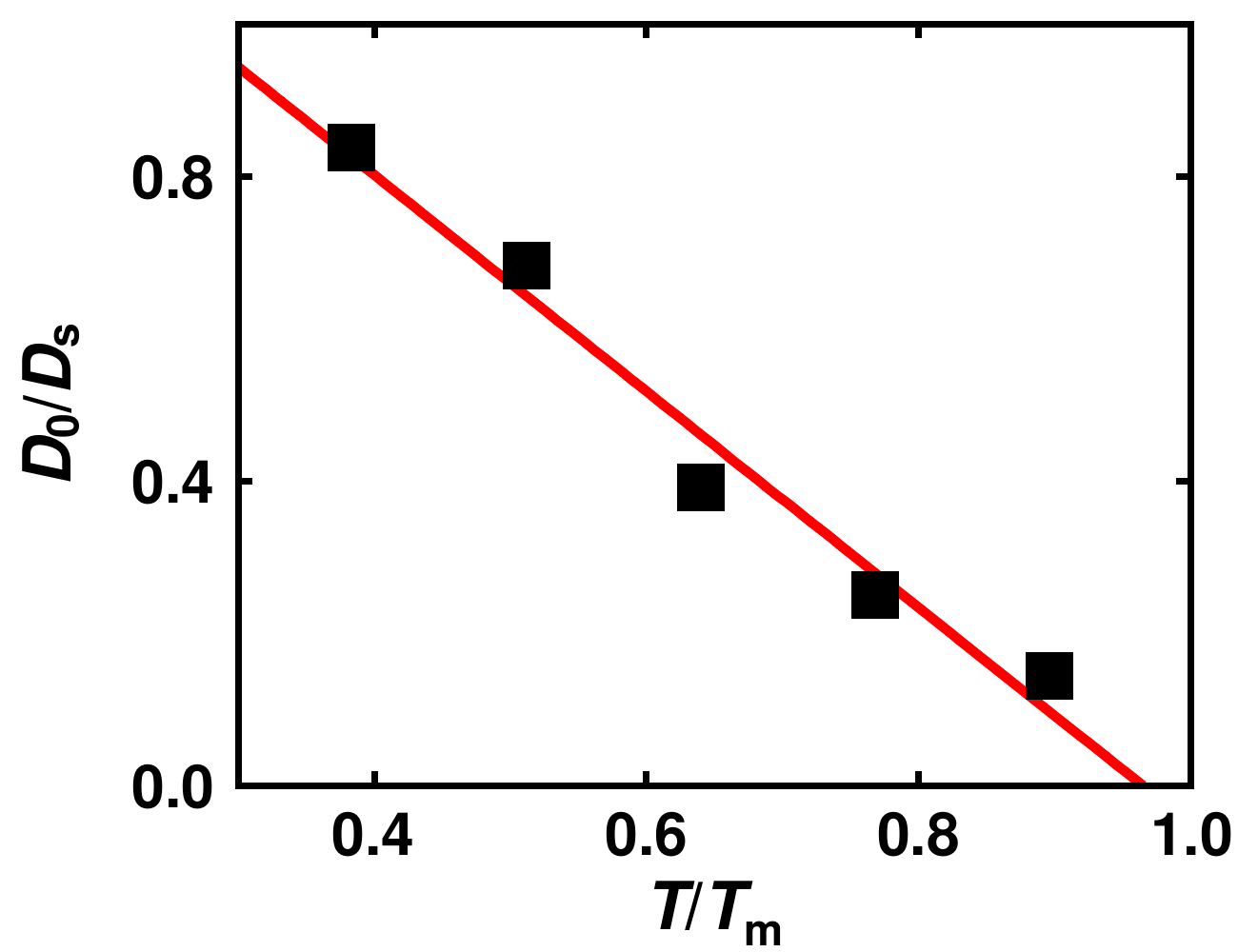}
\caption{\label{DT}Temperature dependence of stagnated grain size $D_\text{s}$ from mesoscale MC simulations (data points) from Ref.~\cite{Holm2010} and a linear fit from \eqref{ktdc}. 
$D_0$ and $T_\text{m}$ are the initial grain size and melting point, respectively. 
}
\end{figure}

The inverse of the critical grain size  is 
\begin{equation}\label{ktdc}
D_\text{KT}^{-1}
= \left(\frac{1}{\epsilon(r_\text{c})}
- \frac{3 k_\text{B} T}{2 Kb^2 w}\right)
\frac{Kb^2}{r_\text{c} \gamma h}.
\end{equation}
Holm and Foiles observed grain growth stagnation at different temperatures in Monte Carlo simulations of polycrystals~\cite{Holm2010}; their data (points in Fig.~\ref{DT}) shows that inverse mean grain size at which stagnation occurs $D_\text{s}^{-1}$ varies with temperature $T$ in an approximately linear function of $T$; as predicted here, \eqref{ktdc}. 
When the grain size exceeds $D_\text{KT}$, the grain will stop growing or shrinking. 
\eqref{ktdc} suggests that, grain growth continues  when $T \ge 2Kb^2w / 3\epsilon k_\text{B} \equiv T_\text{c}$ ($D_\text{KT} \to \infty$).

For nickel (assuming  $b$, $h$, $w$ and $r_\text{c}$ are of the order of one  lattice constant, $\epsilon\approx 1$, and $\gamma\approx 1$~J/m$^2$), we find that $T_\text{c} \sim 18000$~K, which is much higher than the melting point.
This implies that grain growth in polycrystalline nickel should always stagnate at a finite grain size, as observed in  MD simulations~\cite{Holm2010}.
$T_\text{c}$ may decrease substantially upon application of an external stress.

Since $D_\text{KT}$ varies grain-to-grain in a polycrystal, some GBs will show very small mobilities while others will remain mobile. 
As noted by Holm et al.~\cite{Holm2003}, this suggests that abnormal grain growth may readily occur prior to overall grain growth stagnation.

The GB mobility is a tensor, linking both GB shear coupling and migration \cite{Chen2020}.
While the GB migration mobility shows a rapid increase at $T_\text{KT}$, the GB 
sliding coefficient will also increase rapidly at the KT transition temperature. 
This suggests the existence of a GB sliding transition; consistent with the widespread observations of the onset of superplasticity at small grain sizes or high temperature \cite{Edington1976} and intergranular fracture at large grain size and low temperature in many materials \cite{Dowling1999}.

\section*{Discussion}

The theoretical analysis presented above demonstrates that GBs undergo a finite-temperature, Kosterlitz-Thouless, topological phase transition. 
The topological phase transition implies a transition from smooth to rough GBs, a transition from nearly immobile to highly mobile GBs, and a transition from non-sliding to readily sliding GBs.
Because disconnections have dislocation, in addition to step, characters, this transition is topological in nature.
While the step character is associated with the rapid change in GB mobility and GB roughening,  the dislocation character is associated with the onset of GB sliding at  $T_\text{KT}$.

\begin{table*}[t]
\centering
\renewcommand\arraystretch{1.0}
\caption{\label{transitionT}Transition temperatures for thermodynamic GB roughening and sliding and where abrupt changes in GB mobilities are expected for pure step ($\mathbf{b}=\mathbf{0}$), pure dislocation ($h=0$), a single disconnection mode ($\mathbf{b},h$), and multiple disconnection modes ($\mathbf{b}_m,h_m$).  A ``-'' and ``0'' indicate no transition and a transition temperature at $0$ K.  The subscripts and superscripts indicate dimensionality (2d/3d) and pure step ($S$), pure dislocation ($D$), single disconnection  (1), and multiple disconnection  ($M$) modes. $M_{11}$, $M_{12}$ and $M_{22}$ represent  mobilities associated with pure GB migration, shear coupling, and sliding, respectively\cite{Chen2020}. For multiple disconnection modes, entries only represent the lowest temperature of abrupt mobility changes.
}
\begin{tabular}{C{1.0cm}C{0.6cm}C{1.4cm}C{1.4cm}C{1.4cm}C{1.4cm}C{1.4cm}C{1.4cm}C{1.4cm}C{2.0cm}}
\hline
& & \multicolumn{2}{c}{Pure Step} & \multicolumn{2}{c}{Pure Dislocation} & \multicolumn{2}{c}{1 Disconnection Mode} & \multicolumn{2}{c}{Multiple Disconnection Modes} \\
 \cline{3-10}
& & 2d & 3d & 2d & 3d & 2d & 3d & 2d & 3d \\
\hline
\multicolumn{2}{c}{Roughening} & $0$ & $T_3^S$ & - & - & $T_2^1$ & $T_3^1$ & $0$ & $T_{3r}^M$ \\
\multicolumn{2}{c}{Sliding} & - & - & $T_2^D$ & $T_3^D$ & $T_2^1$ & $T_3^1$ & $T_2^M$ & $T_{3s}^M$ \\
\hline
\multirow{3}{*}{Mobility} & $M_{11}$ & - & $T_3^S$ & - & - & $T_2^1$ & $T_3^1$ & $T_2^M$ & $\min(T_{3r}^M,T_{3s}^M)$ \\
& $M_{12}$ & - & - & - & - & $T_2^1$ & $T_3^1$ & $T_2^M$ & $T_{3s}^M$ \\
& $M_{22}$ & - & - & $T_2^D$ & $T_3^D$ & $T_2^1$ & $T_3^1$ & $T_2^M$ & $T_{3s}^M$ \\
\hline
\end{tabular}
\end{table*}

The nature of the dynamic phase transition at GBs depends on disconnection character $\{\mathbf{b},h\}$ and dimensionality (2d or 3d), as summarized in Table~\ref{transitionT}. 
For a pure step ($\mathbf{b}=0$), the transition occurs at $T=0$ in 2d and at finite $T$ in 3d \cite{Swendsen1977}. 
Since this disconnection has $\mathbf{b}=0$, such a transition leads to roughening and an increase in the GB migration mobility, but not to sliding.
For a pure dislocation ($h=0$), the transition occurs at finite temperature in both 2d and 3d. 
Since this disconnection has no associated step, such a transition leads to GB sliding, but not roughening.
For GB dynamics with a single disconnection mode (finite $\mathbf{b}$ and $h$),  the phase transition leads to roughening,  sliding, and a change in all types of mobilities at the same finite temperature $T=T_\text{KT}=T^1_d$.

While we do not explicitly consider multiple disconnection modes here, we expect that (i) the GB roughening will occur at $T=0$ in 2d and at finite temperature ($T_{3r}^M$) in 3d (since pure step modes are always possible) and (ii) a sliding transition at finite temperature in both 2d ($T_{2s}^M$) and 3d ($T_{3s}^M$). 
In the multi-mode case, the mobilities will change abruptly at the topological transitions associated with both the thermodynamic roughening and sliding transitions.
In 2d, the GB sliding transition temperature is associated with the smallest, nonzero Burgers vector.
Above this temperature, all elastic interactions are screened ($\epsilon\to\infty$) and no additional KT transitions will occur; i.e., there is only one sliding transition in 2d (even when multiple disconnection modes are active).
In 3d, two sliding transitions are possible since not all $\mathbf{b}$ are parallel (i.e., the GB is two-dimensional).

Several researchers have demonstrated that grain growth in pure materials often stagnates at a finite grain size \cite{Holm2010}; stagnation is also seen as a pre-requisite to abnormal grain growth (a small set of grains grow to be much larger than the mean grain size) \cite{Holm2003}.
Both stagnation and abnormal grain growth may further or hinder achievement of targeted material properties. 
The presented observations suggest that grain growth stagnation is associated with the difficulty of  disconnection formation/migration below the GB transition temperature ($T<T_\text{KT}$; see Eq.~\ref{ktdc} and Fig. ~\ref{kmcresults}).
This is clear in our 2d simulations, where the GB mobility increases rapidly above $T_\text{KT}$ (Fig.~\ref{kmcresults}g) whereas the roughening temperature is 0 K (\textit{cf.} Fig. ~\ref{kmcresults}a; the 2d multi-mode cases in Table~\ref{transitionT}.)

Ample evidences (experiments, simulations and theories) demonstrate that many aspects of GB dynamics are associated with the formation and motion of  disconnections~\cite{Han2018}.
We presented mean-field and renormalization group theory results and kinetic Monte Carlo evidence for a finite-temperature, disconnection unbinding phase transition (of the Kosterlitz-Thouless type) in GBs.
This disconnection unbinding phase transition is characterized by a finite-temperature transition in GB migration, roughening and sliding.
Associated with these are abrupt changes in the activation energies for the mobilities associated with GB migration and GB sliding at $T_\text{KT}$.
These results provided a unified view of widely-observed feature of grain growth stagnation, abnormal grain, and superplasticity. 
Finally, we note that while other types of GB phase transitions (e.g., first-order structural phase transitions or other transitions in the atomic-scale structure of the GB) may occur and affect GB properties, the disconnection-based KT transition theory gives a unified vision of a wide range of physical phenomena and testable predictions of how these depend on both temperature and grain size.

\section*{Methods}
The model adopted for the kMC simulations is shown schematically in Fig.~\ref{KMCmodel}b. 
The kMC algorithm is as follows. 
\begin{itemize}
\item[(i)]
Initialize the GB configuration $(u_i, y_i)$ at each site $i$. 

\item[(ii)]
List all possible events. 
The event $(im)$ represents the nucleation of a disconnection pair of mode $m$ at site $i$: $(u_i,y_i) \to (u_i + b_m, y_i + h_m) \equiv (u_i^+, y_i^+)$. 

\item[(iii)]
Calculate the energy barriers for event $(im)$:
\begin{equation}
\Delta E_{im}
= (\Delta E_{im}^\text{c} + W_{im}^\text{I} + W_m^\text{E})/2 + E_m^*, \nonumber
\end{equation}
where $\Delta E_{im}^\text{c}$ is the change in core energy, $W_{im}^\text{I}$ is the work done by the stress $\tau_i$, $W_{im}^\text{E}$ is the external work done by the driving forces, and $E_m^*$ is the disconnection glide barrier. 
See Ref.~\cite{Chen2020a} for the formula of each term. 

\item[(iv)]
The rate associated with event $(im)$ is 
$\lambda_{im}
= \omega \exp\left(-\beta\Delta E_{im}\right)$, 
where $\omega$ is the attempt frequency. 
The ``activity'' of the system is $\Lambda = \sum_{i,m} \lambda_{im}$. 

\item[(v)]
Randomly choose an event with probability $p_{im} = \lambda_{im}/\Lambda$. 
Suppose that the selected event is $(i'm')$.

\item[(vi)]
Advance the clock by $\Delta t = \Lambda^{-1}\ln(\eta^{-1})$, where $\eta \in (0,1]$ is a random number. 

\item[(vii)]
Update the state at site $i'$ as $u_{i'} := u_{i'} + b_{m'}$, $y_{i'} := y_{i'} + h_{m'}$ and the stress at each site. 
Return to Step (iii). 
\end{itemize}
In the simulations shown, we set $\zeta=0$, $\gamma=0.1$, $E^\ast =0.1\gamma(|b|+|h|)$.
For the simulation of a GB in thermal equilibrium, we equilibrate the GB at each temperature in the absence of a driving force. 
All data is contained in the main text.

For simplicity, we employ the dimensionless variables: $\tilde{\gamma}=\gamma/2\pi K\delta$, $\tilde{h} = h/\delta$, $\tilde{b} = b/\delta$, $\tilde{y} = y/\delta$, $\tilde{t} = t\omega$, $\tilde{M} = 2\pi KM/\omega\delta$, $\tilde{T} = k_\text{B} T/2\pi K\delta^2w$, $\tilde{\psi}=\psi/2\pi K$, and $\tilde{\tau} = \tau/2\pi K$. 
We drop the ``tilde'' in   \textit{Kinetic Monte Carlo Simulations}  for notational simplicity.

\acknow{The research contribution of  K.C., J.H.,  and D.J.S. (in part) was sponsored by the Army Research Office and was accomplished under Grant Number W911NF-19-1-0263. The views and conclusions contained in this document are those of the authors and should not be interpreted as representing the official policies, either expressed or implied, of the Army Research Office or the U.S. Government. The U.S. Government is authorized to reproduce and distribute reprints for Government purposes notwithstanding any copyright notation herein.  D.J.S. also acknowledges support from the Hong Kong Research Grants Council General Research Fund 11211019.
J.H. also acknowledges support from CityU Strategic Research Grant for unfunded GRF/ECS (SRG-Fd) 7005466, Hong Kong.}

\showacknow{}

\bibliography{mybib}

 \end{document}


\title{{\sc{supplementary material}}\\
Grain-Boundary Topological Phase Transitions}

\author{Kongtao Chen}
\affiliation{Department of Materials Science and Engineering, University of Pennsylvania, Philadelphia, PA 19104 USA}
\author{David J. Srolovitz}
\affiliation{Department of Materials Science and Engineering, City University of Hong Kong, Kowloon, Hong Kong, China}  
\author{Jian Han}
\email{corresponding author: jianhan@cityu.edu.hk}
\affiliation{Department of Materials Science and Engineering, City University of Hong Kong, Kowloon, Hong Kong, China}

\date{\today}

\maketitle

\section{Grain Boundary Roughness}

Consider a (1d) tilt grain boundary (GB) separating a pair of (2d) grains. 
In this situation, disconnections nucleate and migrate along the $x$-axis.
For simplicity, we assume that there is only one disconnection mode $(b, h)$ activated on the GB. 
When a disconnection pair nucleates, a hypothetical ``string'' is defined between them. 
The length of the string is the separation of the two disconnections $r$. 
The sign of the string is the same as the sign of $h$. 
We assume that at position $x$, the number of positive strings is $v_+(x)$, and the number of negative strings is $v_-(x)$; thus, the GB height at $x$ is $y(x) = h[v_+(x)-v_-(x)]$ and the total string number is $v(x)=v_+(x)+v_-(x)$. 

Assume that when the GB is at equilibrium at a particular temperature, $n$ is constant at any point along the GB and at any time. 
At an arbitrary point, there are $n$ strings; the probability for them being positive is $1/2$, i.e., 
\begin{equation}\label{nT1}
\langle v_+\rangle = \frac{v}{2}. 
\end{equation}
The average squared number of positive strings is 
\begin{equation}\label{nT2}
\langle v_+^2\rangle 
= \sum_{i=0}^v i^2 P(v_+ = i)
= \sum_{i=0}^v
\frac{i^2}{2^v} 
{n \choose i} 
= \sum_{i=0}^v
\frac{i^2}{2^v} 
\frac{v!}{i! (v - i)!}
= \frac{v^2 + v}{4}. 
\end{equation}
According to the definition of GB roughness, the squared roughness is 
\begin{align}\label{HT}
\sigma_y^2 
&= \langle y^2\rangle 
= h^2 \langle (v_+-v_-)^2\rangle 
\nonumber\\
&= h^2 \langle (2v_+ - v)^2\rangle 
= h^2 \langle 4v_+^2 + v^2 - 4v_+ v\rangle 
\nonumber\\
&= h^2 \left(
4\langle v_+^2\rangle + v^2 - 4v\langle v_+\rangle
\right)
= h^2 v. 
\end{align}
So, the GB roughness is proportional to the total number of strings per point (i.e., the string density). 

Assume that there are many disconnection dipoles on the GB, with separation $r$ and the distance between dipoles $L$.
Then, the density of disconnection dipoles is $\langle L\rangle^{-1}$. 
Each pair contributes a string of length $\langle r \rangle$.
Thus, the average total string number at one point is 
\begin{equation}\label{nT}
\langle v\rangle 
= \frac{S}{\langle L \rangle}\frac{\langle r \rangle}{S}
=\frac{\langle r\rangle}{\langle L\rangle}
\approx
\left\langle\frac{r^2}{L^2}\right\rangle^{1/2}
= \frac{e^{-2\beta E_\text{c}}}{(2\beta Kb^2 - 3)(2\beta K^2 - 1)},  
\end{equation}
where $S$ is the total length of the 1d GB. 
So, $v$ and thus $\sigma_y^2$ diverge at the critical temperature 
\begin{equation}\label{kttc0}
T_\text{KT}=2Kb^2/3k_\text{B}. 
\end{equation}
Hence, for a 1d GB, roughening transition occurs at $T_\text{KT}$, i.e., the KT transition temperature. 

When $T<T_\text{KT}$, both $r$ and $L$ are finite and as long as $S \gg r$ and $l$, the roughness should be nearly size-independent.
When $T_\text{KT}<T<3 T_\text{KT}$, however, $r\rightarrow\infty$ and the finite size effect is strong.
In this case, the average string length contributed from each pair is $S$ instead of $\langle r \rangle$ in a finite size simulation.
Replacing $\langle r \rangle$ with $S$ in Eq.\eqref{HT} and \eqref{nT}, we have 
\begin{equation}\label{HN1}
\begin{split}
\sigma_y^2 &= \frac{Sh^2e^{-2\beta E_c}}{\delta (2\beta Kb^2-1)}\propto S.
\end{split}
\end{equation}
When $T>3 T_\text{KT}$, $\langle L \rangle < \delta$. This introduces another size effect such that $\langle L \rangle$ should be replaced with $\delta$ in Eq.\eqref{HT} and \eqref{nT},
\begin{equation}\label{HN2}
\begin{split}
\sigma_y^2 &= \frac{Sh^2}{\delta}\propto S.
\end{split}
\end{equation}